\documentclass[5p,twocolumn]{elsarticle}

\usepackage[utf8]{inputenc}
\usepackage{graphicx}
\usepackage{verbatim}
\usepackage[T1]{fontenc}
\usepackage[english]{babel}
\usepackage{bm}
\usepackage{enumerate}
\usepackage{amsmath}
\usepackage{amssymb}
\usepackage{amsbsy}
\usepackage{amsthm}
\usepackage{bbold}
\usepackage{soul}
\usepackage{color}
\usepackage{subfigure}
\usepackage[shortcuts]{extdash}
\usepackage{upgreek}
\usepackage{mathtools}
\usepackage[colorlinks=true, citecolor=blue]{hyperref}
\usepackage[normalem]{ulem}
\usepackage[dvipsnames]{xcolor}
\usepackage{mathrsfs}
\usepackage{amstext}
\usepackage{wasysym}
\usepackage{esint}

\begin{document}
	\global\long\def\a{\alpha}%
	\global\long\def\b{\beta}%
	\global\long\def\c{\chi}%
	\global\long\def\d{\delta}%
	\global\long\def\e{\epsilon}%
	\global\long\def\f{\phi}%
	\global\long\def\g{\gamma}%
	\global\long\def\h{\eta}%
	\global\long\def\i{\iota}%
	\global\long\def\k{\kappa}%
	\global\long\def\l{\lambda}%
	\global\long\def\m{\mu}%
	\global\long\def\n{\nu}%
	\global\long\def\o{\omega}%
	\global\long\def\p{\pi}%
	\global\long\def\q{\theta}%
	\global\long\def\r{\rho}%
	\global\long\def\s{\sigma}%
	\global\long\def\t{\tau}%
	\global\long\def\u{\upsilon}%
	\global\long\def\x{\xi}%
	\global\long\def\y{\psi}%
	\global\long\def\z{\zeta}%
	
	\global\long\def\ve{\varepsilon}%
	\global\long\def\vf{\varphi}%
	\global\long\def\vs{\varsigma}%
	\global\long\def\vq{\vartheta}%
	
	\global\long\def\D{\Delta}%
	\global\long\def\F{\Phi}%
	\global\long\def\G{\Gamma}%
	\global\long\def\L{\Lambda}%
	\global\long\def\Q{\Theta}%
	\global\long\def\S{\Sigma}%
	\global\long\def\U{\Upsilon}%
	\global\long\def\W{\Omega}%
	\global\long\def\X{\Xi}%
	\global\long\def\Y{\Psi}%
	
	\global\long\def\6{\partial}%
	\global\long\def\8{\infty}%
	\global\long\def\j{\int}%
	\global\long\def\w{}%
	\global\long\def\R{\Rightarrow}%
	\global\long\def\*{\times}%
	\global\long\def\={\equiv}%
	\global\long\def\.{\cdot}%
	
	\global\long\def\cA{\mathcal{A}}%
	\global\long\def\cD{\mathcal{D}}%
	\global\long\def\cF{\mathscr{\mathcal{F}}}%
	\global\long\def\cH{\mathcal{H}}%
	\global\long\def\cL{\mathcal{L}}%
	\global\long\def\cJ{\mathcal{J}}%
	\global\long\def\cO{\mathcal{O}}%
	\global\long\def\cP{\mathcal{P}}%
	\global\long\def\cQ{\mathcal{Q}}%
	\global\long\def\cY{\mathcal{Y}}%
	
	\global\long\def\sA{\mathscr{A}}%
	\global\long\def\sD{\mathscr{D}}%
	\global\long\def\sF{\mathscr{F}}%
	\global\long\def\sH{\mathscr{H}}%
	\global\long\def\sL{\mathscr{L}}%
	\global\long\def\sJ{\mathscr{J}}%
	\global\long\def\sO{\mathscr{O}}%
	\global\long\def\sP{\mathscr{P}}%
	\global\long\def\sQ{\mathscr{Q}}%
	\global\long\def\sY{\mathscr{Y}}%
	
	\global\long\def\na{\nabla}%
	\global\long\def\cd{\cdots}%
	\global\long\def\da{\dagger}%
	\global\long\def\ot{\otimes}%
	\global\long\def\we{\wedge}%
	\global\long\def\qu{\quad}%
	
	\global\long\def\fL{\mathfrak{L}}%
	\global\long\def\md{\mathrm{d}}%
	\global\long\def\re{\mathrm{Re}}%
	\global\long\def\im{\mathrm{Im}}%
	\global\long\def\hb{\hbar}%
	
\title{The topological RN-AdS black holes cannot be overcharged by the new version of gedanken experiment}

\author[1]{Yong-Ming Huang}
\ead{huangyongming15@mails.ucas.ac.cn}
\author[1,2]{Yu Tian\corref{cor1}}
\ead{ytian@ucas.ac.cn}
\cortext[cor1]{Corresponding author}
\author[3,4,5]{Xiao-Ning Wu\corref{cor1}}
\ead{wuxn@amss.ac.cn}
\author[6]{Hongbao Zhang\corref{cor1}}
\ead{hongbaozhang@bnu.edu.cn}
\address[1]{School of Physical Sciences, University of Chinese Academy of Sciences, Beijing 100049, China}
\address[2]{Institute of Theoretical Physics, Chinese Academy of Sciences, Beijing 100190, China}
\address[3]{Institute of Mathematics, Academy of Mathematics and System Science, Chinese Academy of Sciences, Beijing 100190, China}
\address[4]{Hua Loo-Keng Key Laboratory of Mathematics, Chinese Academy of Sciences, Beijing 100190, China}
\address[5]{School of Mathematical Sciences, University of Chinese Academy of Sciences, Beijing 100049, China}
\address[6]{Department of Physics, Beijing Normal University, Beijing 100875, China}
\date{\today}

\begin{abstract}
    In this paper, we test the weak cosmic censorship conjecture (WCCC) for $n\geq4$ dimensional nearly extremal RN-AdS black holes with non-trivial topologies, namely plane and hyperbola, using the new version of gedanken experiment proposed by Sorce and Wald. Provided that the non-electromagnetic part of the stress tensor of matter fields satisfies the null energy condition and the linear stability condition holds, we find that the black holes cannot be overcharged under the second-order perturbation approximation, which includes the self-force and finite-size effects. As a result, we conclude that the violation of Hubeny type never occurs and the WCCC holds for the topological RN-AdS black hole.
\end{abstract}


\maketitle

\section{Introduction}
The inevitability of singularities is guaranteed by the singularity theorem proposed by Penrose and Hawking \cite{Penrose:1964wq,Hawking:1970zqf}. The existence of spacetime singularity poses a problem for the general relativity, in the sense that the theory does not apply to the spacetime region close to the singularity. To resolve the problem, Penrose proposed the weak cosmic censorship conjecture \cite{Penrose:1969pc}, stating that the singularities must be hidden from an observer at infinity by the event horizon of a black hole. Due to the difficulty of proving this conjecture universally, researchers hope to test the validity of this conjecture by considering physically reasonable processes. In this regard, Wald first proposed a gedanken experiment to see whether an extremal Kerr-Newman black hole can be destroyed by throwing a test particle into it \cite{wald1974gedanken}. As a result, he showed that the test particle that would cause the black hole to be destroyed cannot be captured, in support of WCCC. But nonetheless, Hubeny found that a near-extremal RN black hole can be destroyed if the infalling matter's parameters are chosen appropriately \cite{Hubeny:1998ga}, and some subsequent studies also claimed to have succeeded in overspining  Kerr black holes \cite{Matsas:2007bj,Jacobson:2009kt} or Kerr-Newman black holes \cite{Saa:2011wq,Gao:2012ca}. However, the back-reaction effects are not taken into account in the above investigations \cite{Hubeny:1998ga,Hod:2002pm,Hod:2008zza,Barausse:2010ka,Barausse:2011vx,Zimmerman:2012zu,Colleoni:2015afa,Colleoni:2015ena}. Thus, if we want to make certain that these potential violations occur, we need a more comprehensive analysis of the back-reaction effects in these processes, which includes but is not limited to self-force and finite-size effects.

For this purpose, Sorce and Wald have recently proposed a new gedanken experiment to test WCCC \cite{Sorce:2017dst}. Rather than analyze the motion of test particles to determine whether or not they fall into black holes, they assume all charged matter that is initially present is absorbed by the black holes. Assume that the non-electromagnetic part of the stress tensor of matter fields satisfies the null energy condition and the linear stability condition holds, they estimate the second-order correction to the black hole parameters, where the self-force and finite size effects are automatically taken into account. As a result, they show that the event horizon of nearly extremal Kerr-Newman black holes cannot be destroyed under the second-order perturbation approximation, although the black holes can be destroyed at the linear order. This demonstrates the back-reaction effects protect the validity of WCCC. With this new gedanken experiment, numerous physical processes involved in the investigation of WCCC have been examined or reexamined \cite{An:2017phb,Ge:2017vun,He:2019mqy,Jiang:2019ige,Jiang:2019vww,Chen:2019nhv,Ding:2020zgg,Jiang:2019soz,Jiang:2020mws,Jiang:2020xow,Jiang:2020btc,Jiang:2020alh,Wang:2020vpn,Zhang:2020txy,Wang:2019bml,Li:2020smq,Qu:2021hxh,Sang:2021xqj,Shaymatov:2020wtj,Wang:2021kcq}, indicating that WCCC holds.

In \cite{Zhang:2020txy,Wang:2019bml}, the weak cosmic censorship conjecture for asymptotically AdS charged black holes with spherical topology have been investigated. However, the perturbation of matter fields considered therein is restricted to be radial. On the other hand, it is well-known that the presence of a negative cosmological constant allows for more diverse topologies for black holes \cite{Birmingham:1998nr,Mann:1996gj,Mann:1997zn,Vanzo:1997gw,Brill:1997mf}. In particular, the AdS black hole horizon with the  maximal symmetry can not only be spherical, but also be topological, namely planar or hyperbolic. Numerous studies have established that the topological black holes with planar and hyperbolic horizons exhibit new properties that are not observed in spherical black holes \cite{Emparan:1999gf}. With this in mind, we aim to apply the new gedanken experiment in a uniform manner to both spherical and topological charged AdS black holes, where the aforementioned perturbation of matter fields is relaxed to be as general as possible. As a result, the violation of Hubeny type never occurs in topological case as in spherical case.

The organization of the paper is as follows. In Section II, we study the perturbation of RN-AdS black holes under the matter field, where the spherical, planar and hyperbolic are treated in a uniform manner. In Section III, based on the null energy condition and linear stability condition, we deduce the first and second-order perturbation inequalities by using Iyer-Wald formalism. In Section III, we test WCCC for the RN-AdS black holes. Section IV includes some concluding remarks.

\section{The perturbed RN-AdS black holes}

Let us start with $n$-dimensional
Einstein-Maxwell gravity with a negative cosmological constant, whose Lagrangian
is written as 
\begin{equation}
L=\frac{\e}{16\pi}\left(R-2\L-F_{ab}F^{ab}\right)+L_{matter},
\end{equation}
where $\e$ is the volume element, $R$ is the Ricci scalar, $F=dA$
is the electromagnetic field strength, with $A$ is the vector potential of
electromagnetic field, $\L$ is the negative cosmological constant
and $L_{matter}$ denotes the Lagrangian of extra matter fields. The
equations of motion (EOM) of this theory can be written as
\begin{eqnarray}
8\pi T_{ab} & = & G_{ab}-\L g_{ab}-8\pi T_{ab}^{EM},\label{eq:1}\\
\nabla_{a}F^{ab} & = & 4\pi j^{b},\nonumber 
\end{eqnarray}
with
\begin{equation*}
	T_{ab}^{EM} =  \text{\ensuremath{\frac{1}{4\pi}\left(F_{ac}F_{b}^{\text{}c}-\frac{1}{4}g_{ab}F_{cd}F^{cd}\right)},}
\end{equation*}
where $T_{ab}$ corresponds to the non-electromagnetic part of the
stress tensor and $j^{a}$ corresponds to the electromagnetic current.

When the extra matter fields  vanish, the theory yields a sequence
of maximally symmetric black hole solutions, denoted by
\begin{eqnarray}
ds^{2} & = & -f(r)d\n^{2}+2drd\n+r^{2}d\Omega_{n-2}^{(k)2},\label{eq:2}\\
A_{\n} & = & -\frac{4\pi Q}{\left(n-3\right)\text{\ensuremath{\Omega_{n-2}^{(k)}r^{n-3}}}}d\n.\nonumber 
\end{eqnarray}
Here $k$ indicates the spatial curvature
of the black hole, and
\begin{eqnarray}
f\left(r\right) & = & k-\frac{p}{r^{n-3}}+\frac{q^{2}}{r^{2n-6}}+\frac{r^{2}}{l^{2}},\label{eq:3}\\
d\Omega_{n-2}^{(k)2} & = & \hat{g}_{ij}^{\left(k\right)}\left(x\right)dx^{i}dx^{j},\nonumber 
\end{eqnarray}
where $l$ is the AdS radius. Specifically, $k=\left\{ 1,0,-1\right\} $ corresponds to spherical, planar and hyperbolic horizon topologies, respectively, and $\Omega_{n-2}^{(1,0,-1)}$ is the volume of $\left(n-2\right)$
dimensional ``unit'' sphere, plane or hyperbola, respectively, where
the latter two (non-compact) have been compactified properly \cite{Tian:2014goa,Tian:2018hlw}.
Furthermore, the parameters $p$ and $q$ are related to the mass and
the charge of the black hole as
\[
p=\frac{16\pi M}{\left(n-2\right)\Omega_{n-2}^{(k)}},\qquad q=\frac{4\pi}{\Omega_{n-2}^{(k)}}\sqrt{\frac{2}{\left(n-2\right)\left(n-3\right)}}Q.
\]

Following that, we consider the case $f\left(r\right)$ has at least two roots, and we refer
to the maximal root of $f\left(r\right)$ as $r_{h}$, which corresponds to the black hole event horizon. Then the temperature,
area and electric potential of the event horizon are given by 
\begin{equation*}
T=\frac{f'\left(r_{h}\right)}{4\pi},\;
\cA=\text{\ensuremath{\Omega_{n-2}^{(k)}}}r_{h}^{n-2},\;
\F_{\cH}=\frac{4\pi Q}{\left(n-3\right)\text{\ensuremath{\Omega_{n-2}^{(k)}r_{h}^{n-3}}}}.
\end{equation*}
Furthermore, when $f\left(r\right)$ has no root, the solutions
describe naked singularities. 

With the new gedanken experiment, we assume that the background
spacetime is linearly stable to perturbations, i.e., the linear perturbed RN-AdS
black hole will evolve into another RN-AdS black hole at sufficiently
late times, then the late spacetime geometry will be generally
written as
\begin{equation}
ds^{2}=-f\left(r,\l\right)d\n^{2}+2drd\n+r^{2}d\Omega_{n-2}^{(k)2},\label{eq:4}
\end{equation} where
\begin{equation*}
	f\left(r,\l\right) =  k-\frac{p\left(\l\right)}{r^{n-3}}+\frac{q^{2}\left(\l\right)}{r^{2n-6}}+\frac{r^{2}}{l^{2}}.
\end{equation*}

\section{Null energy conditions and perturbation inequalities}

In this section, we will use the Iyer-Wald formalism \cite{Iyer:1994ys,Iyer:1995kg} to derive
perturbation inequalities. Consider a $n$-dimensional Einstein-Maxwell
theory with a negative cosmological constant, whose $n$-form Lagrangian is written
as
\begin{equation}
L=\frac{\e}{16\pi}\left(R-2\L-F_{ab}F^{ab}\right).\label{eq:6}
\end{equation}
Then we focus on such a one-parameter family $\f(\l)$  that satisfies the EOM \eqref{eq:1} and describes the whole perturbation process, where $\f$ denotes $(g_{ab},A_{a})$ and extra matter fields. 

The first-order variation of the Lagrangian gives
\begin{equation}
\d L = E_{\f}\d\f+d\Theta\left(\f,\d\f\right),\label{eq:7}
\end{equation}
\begin{equation}
\begin{split}
E_{\f}\d\f & = -\e\left(\frac{1}{2}T^{ab}\d g_{ab}+j^{a}\d A_{a}\right),\nonumber \\
\Theta\left(\f,\d\f\right) & =\Theta^{GR}\left(\f,\d\f\right)+\Theta^{EM}\left(\f,\d\f\right),\nonumber \\
\Theta_{a_{2}\cdot\cdot\cdot a_{n}}^{GR}\left(\f,\d\f\right) & = \frac{1}{16\pi}\e_{da_{2}\cdot\cdot a_{n}}g^{de}g^{fh}\left(\na_{h}\d g_{ef}-\na_{e}\d g_{fh}\right),\nonumber \\
\Theta_{a_{2}\cdot\cdot\cdot a_{n}}^{EM}\left(\f,\d\f\right) & = -\frac{1}{4\pi}\e_{da_{2}\cdot\cdot a_{n}}F^{de}\d A_{e},\nonumber 
\end{split}
\end{equation}
where $E_{\f}=0$ is the EOM of this theory and $\Theta$ is called
symplectic potential. 

The $\left(n-1\right)$-form symplectic current then is defined as
\begin{equation}
\o\left(\f,\d_{1}\f,\d_{2}\f\right)=\d_{1}\Theta\left(\f,\d_{2}\f\right)-\d_{2}\Theta\left(\f,\d_{1}\f\right).
\end{equation}

The Noether current, which is associated with arbitrary vector field
$\c^{a}$, is defined as
\begin{equation}
J_{\c}=  \Theta\left(\f,\cL_{\c}\f\right)-\i_{\c}L\left(\f\right),
\label{eq:8}\end{equation} where $\i_{\c}L\left(\f\right)$ denotes the vector $\c^{a}$ contracting
with the first index of $n$-form Lagrangian. A simple calculation
which combines (\ref{eq:7}) and (\ref{eq:8}) can show that the exterior
derivative of $J_{\c}\left(\f\right)$ gives 
\begin{equation}
dJ_{\c}= -E_{\f}\cL_{\c}\f.\label{eq:9}
\end{equation}
When the EOM are satisfied, (\ref{eq:9}) reduces to $dJ_{\c}=0$,
implying the Noether current defined by (\ref{eq:8}) satisfies the conserved equation. 

On the other hand, it was shown in \cite{Iyer:1995kg} that
the $\left(n-1\right)$-form Noether current can also be written as
\begin{eqnarray}
J_{\c} & = & C_{\c}+dQ_{\c},\label{eq:10}\\
\left(C_{\c}\right)_{a_{2}\cdot\cdot a_{n}} & = & \c^{a}\e_{ba_{2}\cdot\cdot a_{n}}\left(T_{a}^{b}+A_{a}j^{b}\right),\nonumber \\
Q_{\c} & = & Q_{\c}^{GR}+Q_{\c}^{EM},\nonumber \\
\left(Q_{\c}^{GR}\right)_{a_{3}\cdot\cdot\cdot a_{n}} & = & -\frac{1}{16\pi}\e_{dea_{3}\cdot\cdot a_{n}}\nabla^{d}\c^{e},\nonumber \\
\left(Q_{\c}^{EM}\right)_{a_{3}\cdot\cdot\cdot a_{n}} & = & -\frac{1}{8\pi}\e_{dea_{3}\cdot\cdot a_{n}}F^{de}\c^{c}A_{c},\nonumber 
\end{eqnarray} where $Q_{\c}$ is called the Noether charge and $C_{\c}$ describes
the constraints of the theory. 

Comparing the variation of (\ref{eq:8})
and (\ref{eq:10}), we obtain the first-order variational identity
\begin{equation}
d\left(\d Q_{\c}-\c\cdot\Theta\left(\f,\d\f\right)\right)=\o\left(\f,\d\f,\cL_{\c}\f\right)-\c\cdotp E\d\f-\d C_{\c}.\label{eq:11}
\end{equation}
Varying (\ref{eq:11}) gives the second order variational identity
\begin{equation}
\begin{split}
&d\left(\d^{2}Q_{\c}-\c\cdot\d\Theta\left(\f,\d\f\right)\right)  \\
=\quad &\o\left(\f,\d\f,\cL_{\c}\d\f\right)-\c\cdotp\d(E\d\f) -\d^{2}C_{\c},
\end{split} \label{eq:12}
\end{equation} where we have assumed that $\c^{a}$ is a symmetry of $\f$, i.e, $\cL_{\c}\f=0$.
\begin{figure}
	\includegraphics[scale=0.66]{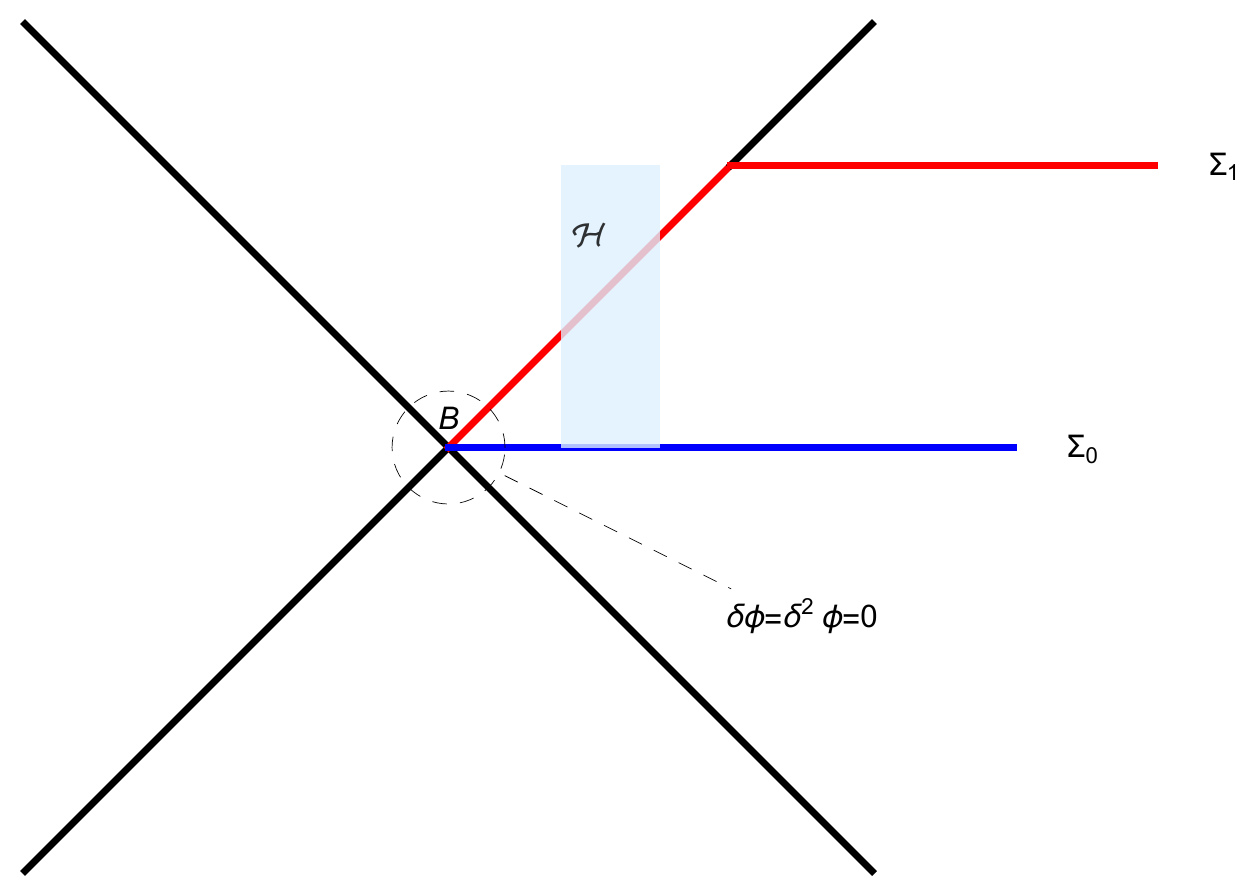}\caption{The shaded region describes the process that the charged matter is absorbed entirely by the black hole. Around the bifurcation surface B, the initial perturbation vanishes.}
\end{figure}

In order to test WCCC, we
assume that all the charged matter enters the black hole via a finite
portion of the future event horizon. Based on this, we can always
choose a hypersurface $\S=\cH\cup\S_{1}$ such that it begins at
the unperturbed bifurcation surface $B$ and continues up
the portion of the horizon $\cH$ until the region where the matter field vanishes at
late times, then become as $\S_{1}$ toward the spatial infinity, see Figure 1. In this framework, we consider the Noether current that is induced
by the killing vector of the background spacetime $\x^{a}=\left(\6_{\n}\right)^{a}$
and work in a gauge where $\cL_{\x}A_{a}=0$, then the integral of
(\ref{eq:11}) on the Cauthy surface of our choice can be written
as
\begin{equation}
\int_{\6\S}\left(\d Q_{\x}-\x\cdot\Theta\left(\f,\d\f\right)\right)+\int_{\S}\x\cdotp E\d\f+\int_{\S}\d C_{\x}=0.\label{eq:13}
\end{equation}
For the first term of left-hand side of (\ref{eq:13}), using the
explicit metric expression (\ref{eq:4}), one can obtain
\begin{equation}
\begin{split}
	&\int_{\6\S}\left(\d Q_{\x}-\x\cdot\Theta\left(\f,\d\f\right)\right) \\
	= & \int_{S_{\8}}\left(\d Q_{\x}-\x\cdot\Theta\left(\f,\d\f\right)\right)-\int_{B}\left(\d Q_{\x}-\x\cdot\Theta\left(\f,\d\f\right)\right)\\
	= & \int_{S_{\8}}\left(\d Q_{\x}^{GR}-\x\cdot\Theta^{GR}\left(\f,\d\f\right)\right)\\
	&+\int_{S_{\8}}\left(\d Q_{\x}^{EM}-\x\cdot\Theta^{EM}\left(\f,\d\f\right)\right)\\
	= & \quad \d M, 
	\end{split}
\end{equation}
where $S_{\8}$ denotes the boundary of $\S_{1}$ at the spatial infinity.

Due to the facts that $\x\cdotp\e=0$ when we restrict on the horizon and there are
no matter fields on $\S_{1}$, it turns out that the second term of
(\ref{eq:13}) vanishes. In the following, we impose a condition
$\x^{a}\d A_{a}|_{\cH}=0$, which can always be achieved through gauge
transformation, then the third term gives
\begin{equation}
\begin{split}
	\int_{\S}\d C_{\x}=& \int_{\cH}\d C_{\x}+\int_{\S_{1}}\d C_{\x}\\
	=&  \d\int_{\cH}\x^{a}\e_{ba_{2}\cdots a_{n}}\left(T_{a}^{b}+A_{a}j^{b}\right)\\
	 =& -\left(\int_{\cH}\x_{a}k_{b}\d T^{ab}\widetilde{\e}_{a_{2}\cdots a_{n}}\right)\\
	 &+\left(\x^{a}A_{a}\right)_{\cH}\d\int_{\cH}\e_{ba_{2}\cdots a_{n}}j^{b}\\
	=&  -\left(\int_{\cH}\x_{a}k_{b}\d T^{ab}\widetilde{\e}_{a_{2}\cdots a_{n}}\right)-\F_{\cH}\d Q,
	\end{split}
\end{equation}
where $k^{a}\propto\x^{a}$ is the future-directed tangent vector
field to the horizon and $\widetilde{\e}_{a_{2}\cdots a_{n}}$denotes
the volume element on the horizon, which is defined as $\e_{ba_{2}\cdots a_{n}}=-nk_{[b}\widetilde{\e}_{a_{2}\cdots a_{n}]}$.

As a result, (\ref{eq:13}) can be rewritten as
\begin{equation}
\d M-\F_{\cH}\d Q=\int_{\cH}\x_{a}k_{b}\d T^{ab}\widetilde{\e}_{a_{2}\cdots a_{n}}.\label{eq:14}
\end{equation}
Assume that the non-electromagnetic stress tensor satisfies the null energy
condition on the horizon, i.e. $\d T^{ab}\x_{a}\x_{b}|_{\cH}\geq0,$
we will obtain an inequality
\begin{equation}
\d M-\F_{\cH}\d Q\geq0,
\end{equation}
which is referred to as the first-order perturbation inequality.

In this paper, we investigate whether WCCC holds for RN-AdS black holes in terms
of the second-order perturbation approximation. For this purpose, we assume
that the first-order perturbation has been done optimally, i.e. $\d T^{ab}\x_{a}\x_{b}|_{\cH}=0$,
which indicates that the non-electromagnetic energy flux associated
with the first-order perturbation through the horizon vanishes. Under
this assumption, one can show that the first-order perturbation of expansion $\d\vq$ vanishes on the horizon $\cH$ \cite{Gao:2001ut,Hollands:2012sf}.

With the optimal condition, we turn to derive the second-order perturbation inequality. Similar to the first-order perturbation, the integral
of (\ref{eq:12}) on $\S$ is written as
\begin{equation}
\begin{split}
&\int_{\6\S}\left[\d^{2}Q_{\x}-\x\cdot\d\Theta\left(\f,\d\f\right)\right]\\
=\quad & \mathscr{E}_{\cH}\left(\f,\d\f\right)+\mathscr{E}_{\S_{1}}\left(\f,\d\f\right)-\int_{\S}\x\cdotp\d(E\d\f)-\int_{\S}\d^{2}C_{\x},
\end{split}
\label{eq:15}
\end{equation}
\begin{equation}
\begin{split}
\mathscr{E}_{\cH}\left(\f,\d\f\right)= &\int_{\cH}\o^{GR}\left(\f,\d\f,\cL_{\x}\d\f\right)\\
&+\int_{\cH}\o^{EM}\left(\f,\d\f,\cL_{\x}\d\f\right),\nonumber \\
\mathscr{E}_{\S_{1}}\left(\f,\d\f\right)= &\int_{\S_{1}}\o\left(\f,\d\f,\cL_{\x}\d\f\right).\nonumber 
\end{split}
\end{equation}

Based on the assumption that the second-order perturbation vanishes
at the bifurcation surface $B$, one can show that the left hand side of (\ref{eq:15})
gives
\begin{equation}
	\int_{\6\S}\left[\d^{2}Q_{\x}-\x\cdot\d\Theta\left(\f,\d\f\right)\right]=\d^{2}M.
\end{equation}

For the gravitational part of the canonical energy $\mathscr{E}_{\cH}\left(\f,\d\f\right)$, we borrow the result from \cite{Hollands:2012sf,Hollands:2014lra,Green:2015kur}, where shows 
\begin{equation}
\int_{\cH}\o^{GR}\left(\f,\d\f,\cL_{\x}\d\f\right)\geq 0.\label{eq:16}
\end{equation}

Furthermore, the electromagnetic part of $\mathscr{E}_{\cH}\left(\f,\d\f\right)$,
whose calculation is similar to \cite{Sorce:2017dst}, can be showed as
\begin{equation}
\begin{split}
	\int_{\cH}\o^{EM}\left(\f,\d\f,\cL_{\x}\d\f\right)&= -\frac{1}{2\pi}\int_{\cH}\x^{c}\d F^{ab}\d F_{cb}\e_{aa_{2}\cdot\cdot\cdot\cdot\cdot a_{n}} \\&=\int_{\cH}\d^{2}T_{ab}^{EM}k^{a}\x^{b}\widetilde{\e}\geq0,\label{eq:16-1}
	\end{split}
\end{equation}
which implies the total flux of electromagnetic energy into the black hole is nonnegative.

Furthermore, it is obvious that the third term of right-hand side of
(\ref{eq:15}) vanishes, and the fourth term gives
\begin{equation}
\begin{split}
\int_{\S}\d^{2}C_{\x}&=\int_{\S_{1}}\d^{2}C_{\x}+\int_{\cH}\d^{2}C_{\x}\\
&=\int_{\cH}\d^{2}C_{\x} \\
& = -\int_{\cH}\widetilde{\e}\left(\d^{2}T^{ab}\x_{a}k_{b}\right)-\F_{\cH}\d^{2}Q. 
\end{split}
\label{eq:17}
\end{equation}

Until now, the only term that has not been obtained yet is the canonical
energy on $\S_{1}$. In order to calculate this term, we consider
another one-parameter family of field configuration $\f^{RA}\left(\l\right)$,
which describes the RN-AdS black hole with parameters given by
\begin{equation}
\begin{split}
	M^{RA}\left(\l\right)&= M+\l\d M,\\
	Q^{RA}\left(\l\right)&= Q+\l\d Q,
	\end{split}
\end{equation}
where $\d M$ and $\d Q$ equal to the corresponding value of our previous first-order perturbation when we set $\d\f^{RA}=\d\f$. For this family, we have $\d^{2}M=\d^{2}Q=\d^{2}C_{\x}=\d E=\mathscr{E}_{\cH}\left(\f,\d\f^{RA}\right)=0$, then (\ref{eq:15}) gives
\begin{equation}
\begin{split}
	\mathscr{E}_{\S_{1}}\left(\f,\d\f\right)&=\mathscr{E}_{\S_{1}}\left(\f,\d\f^{RA}\right) \\
	&= -\int_{B}\left[\d^{2}Q_{\x}-\x\cdot\d\Theta\left(\f,\d\f^{RA}\right)\right]\\
	& = \frac{4\pi}{\left(n-3\right)r_{h}^{n-3}\Omega_{n-2}^{\left(k\right)}}\left(\d Q\right)^{2}.
\end{split}
\end{equation}

Putting these results together, we conclude that (\ref{eq:15}) can result in the following second-order perturbation inequality, which is written as 
\begin{equation}
\d^{2}M-\F_{\cH}\d^{2}Q-\frac{4\pi}{\left(n-3\right)r_{h}^{n-3}\Omega_{n-2}^{\left(k\right)}}\left(\d Q\right)^{2}\geq0,\label{eq:18}
\end{equation} 
where we have assumed that the null energy condition for matter fields
is satisfied, i.e. $\d^{2}T_{ab}\geq0$. 

\section{Gedanken experiment to destroy a nearly extremal RN-AdS black hole}

In this section, we attempt to overcharge a nearly extremal RN-AdS
black hole using the previously mentioned gedanken experiment. All
that we need to do is to determine whether the perturbed geometry at
late times still describes a black hole, which is equivalent to investigating
if the solution of $f\left(r,\l\right)=0$ exists. For simplicity,
we consider the function
\begin{equation}
F\left(r,\l\right)=r^{n-3}f\left(r,\l\right)=kr^{n-3}-p\left(\l\right)+\frac{q^{2}\left(\l\right)}{r^{n-3}}+\frac{r^{n-1}}{l^{2}},
\end{equation} whose the largest root gives the location of the event horizon $r_{h}$.

Moreover, we define
\begin{equation}
\begin{split}
g\left(\l\right)&=F\left(r_{m}\left(\l\right),\l\right)\\
&=kr_{m}^{n-3}\left(\l\right)-p\left(\l\right)+\frac{q^{2}\left(\l\right)}{r_{m}^{n-3}\left(\l\right)}+\frac{r_{m}^{n-1}\left(\l\right)}{l^{2}},
\end{split}
\label{eq:19}
\end{equation} 
where $r=r_{m}\left(\l\right)$ gives the minimal value of $F\left(r,\l\right)$. So we have
\begin{equation}
\6_{r}F\left(r_{m}\left(\l\right),\l\right)=0\label{eq:20}
\end{equation}
and
\begin{equation}
\d\left[\6_{r}F\left(r_{m}\left(\l\right),\l\right)\right]=0,\label{eq:21}
\end{equation} 
which imply that
\begin{equation}
\d r_{m}=\frac{-2\left(3-n\right)r_{m}^{2-n}\left(\l\right)q\left(\l\right)\d q}{\6_{r}^{2}F\left(r_{m}\left(\l\right),\l\right)}.\label{eq:22}
\end{equation} 
Notice that the violation of the black hole happens when $g\left(\l\right)>0$. For convenience, we will refer to $r_{m}(0)$,$r_{h}(0)$ as $r_{m},r_{h}$ respectively in the following.  

Considering (\ref{eq:20}) and (\ref{eq:21}), we expand $g\left(\l\right)$
to the second-order and obtain
\begin{equation}
\begin{split}
g\left(\l\right)&= F\left(r_{m},0\right)+\l\left(-\d p+\frac{2q\d q}{r_{m}^{n-3}}\right) \\
 &+\frac{\l^{2}}{2}\left(-\d^{2}p+\frac{2q\d^{2}q}{r_{m}^{n-3}}+\frac{2\left(\d q\right)^{2}}{r_{m}^{n-3}}+\left(3-n\right)\frac{2q\d q\d r_{m}}{r_{m}^{n-2}}\right)\\
  &+O\left(\l^{3}\right).
\end{split}
\label{eq:23}
\end{equation}

Let us consider a nearly extremal black hole as the background
spacetime, whose event horizon has the relationship
$r_{m}=\left(1-\e\right)r_{h}$, with a small parameter $\e>0$. With above preparation, (\ref{eq:20}) gives
\begin{equation}
	\6_{r}F\left(r_{h},0\right) = \e r_{h}\6_{r}^{2}F\left(r_{h},0\right)+O\left(\e^{2}\right).
\end{equation} 
Thus, for the first term of (\ref{eq:23}), we have
\begin{equation}
\begin{split}
F\left(r_{m},0\right)&=F\left(\left(1-\e\right)r_{h},0\right) \\
&=-\e r_{h}\6_{r}F\left(r_{h},0\right)+\frac{1}{2}\e^{2}r_{h}^{2}\6_{r}^{2}F\left(r_{h},0\right)+O\left(\e^{3}\right) \\
&=-\frac{1}{2}\e^{2}r_{h}^{2}\6_{r}^{2}F\left(r_{h},0\right)+O\left(\e^{3}\right).
\end{split} \label{eq:24}
\end{equation}

Combine (\ref{eq:22}) and (\ref{eq:24}), (\ref{eq:23})
gives
\begin{equation}
\begin{split}
g\left(\l\right)=& -\frac{1}{2}\e^{2}r_{h}^{2}\6_{r}^{2}F\left(r_{h},0\right)+\left(-\d p+\frac{2q\d q}{r_{h}^{n-3}}\right)\l\\
&-\frac{2\left(3-n\right)q\d q}{r_{h}^{n-3}}\e\l +\left(-\d^{2}p+\frac{q\d^{2}q}{r_{h}^{n-3}}+\frac{\left(\d q\right)^{2}}{r_{h}^{n-3}}\right)\l^{2} \\
&-\frac{2}{\6_{r}^{2}F\left(r_{h},0\right)}\left(\left(3-n\right)r_{h}^{2-n}q\d q\right)^{2}\l^{2}\\
 & +O\left(\l^{3},\e^{3},\e\l^{2},\l\e^{2}\right) \\
\leq&-\frac{1}{2}\e^{2}r_{h}^{2}\6_{r}^{2}F\left(r_{h},0\right)-\frac{2\left(3-n\right)q\d q}{r_{h}^{n-3}}\e\l \\
& -\frac{2}{\6_{r}^{2}F\left(r_{h},0\right)}\left(\left(3-n\right)r_{h}^{2-n}q\d q\right)^{2}\l^{2} \\
&+O\left(\l^{3},\e^{3},\e\l^{2},\l\e^{2}\right) \\
= & -\frac{\left(\e r_{h}\6_{r}^{2}F\left(r_{h},0\right)+2\left(3-n\right)r_{h}^{2-n}q\d q\l\right)^{2}}{2\6_{r}^{2}F\left(r_{m},0\right)}\\
&+O\left(\l^{3},\e^{3},\e\l^{2},\l\e^{2}\right).
\end{split}
\label{eq:25}
\end{equation} 
Here we have used the optimal condition and the second-order perturbation
inequality in the second step. In the last step, we have used the
fact that $F\left(r,0\right)$ has the minimal value at $r=r_{m}$,
implying $\6_{r}^{2}F\left(r_{m},0\right)>0$. 

In summary, (\ref{eq:25}) demonstrates that  it is possible to have $g\left(\l\right)>0$ at the linear order level, indicating that an nearly extremal RN-AdS black hole with non-trivial topologies can be overcharged into a naked
singularity. However, it also shows that the nearly extremal black hole cannot be overcharged when the second-order perturbation is taken into account, i.e. the WCCC holds in this case.

\section{Concluding remarks}
In this paper, we have used the new version of gedanken experiment proposed by Sorce and Wald to test the weak cosmic censorship conjecture for nearly extremal RN-AdS black holes, where the topological and spherical cases are treated in a uniform manner. We begin with reviewing the perturbation processes of the RN-AdS black holes. Then, using the Iyer-Wald formalism, we have deduced the first and second-order perturbation inequalities based on the linearly stable condition and the assumption that the non-electromagnetic part of the stress tensor of matter fields satisfies null energy condition. Finally, we have investigated whether the Hubeny type violation occurs under the second-order perturbation approximation and discovered that the perturbed black holes satisfy the black hole condition, implying that WCCC holds not only for the spherical case, but also for the planar and hyperbolic cases.

\section*{Acknowledgements}
  This work is partially supported by the National Natural Science
  Foundation of China with Grant No.11731001, 11875095, 11975235, 12035016 and 12075026.


\end{document}